\documentclass[aps,superscriptaddress,prd,onecolumn,floatfix, nofootinbib,amsmath,amssymb,amsfonts,longbibliography]{revtex4-2}

\usepackage{mathrsfs,amsmath}
\usepackage[pdftex]{graphicx}
\usepackage[dvipsnames]{xcolor}
\usepackage{float}
\usepackage{physics}
\usepackage{comment}
\usepackage[normalem]{ulem}
\usepackage{caption}
\usepackage{subcaption}
\usepackage{tabularx}

   \newcolumntype{C}{>{\centering\arraybackslash}X}
   \newcolumntype{L}{>{\raggedright\arraybackslash}X}
   \newcolumntype{R}{>{\raggedleft\arraybackslash}X}
   
   \usepackage{hhline}

\usepackage{bm}
\usepackage[accsupp]{axessibility}
\usepackage[colorlinks=true,linkcolor=blue,anchorcolor=violet,citecolor=red]{hyperref}
\newcommand{\souteq}[1]{\hbox{}}

\begin{document}

\begin{flushright}
\textbf{Report number: NU-QG-1}
\end{flushright}

\title{Test particle motion around a black hole \\dressed with a spherically symmetric stationary fluid}

\author{Ariadna Uxue Palomino Ylla}\email{palomino.ylla.ariadna.uxue.e8@s.mail.nagoya-u.ac.jp}
\affiliation{Division of Science, Graduate School of Science, Nagoya University, Nagoya 464-8602, Japan}

\author{Yasutaka Koga}\email{yasutaka.koga@yukawa.kyoto-u.ac.jp}
\affiliation{Department of Information and Computer Science, Osaka Institute of Technology, Hirakata 573-0196, Japan}
\affiliation{Yukawa Institute for Theoretical Physics,
Kyoto University, Kyoto 606-8502, Japan}

\author{Chul-Moon Yoo}\email{yoo.chulmoon.k6@f.mail.nagoya-u.ac.jp}
\affiliation{Division of Science, Graduate School of Science, Nagoya University, Nagoya 464-8602, Japan}
\affiliation{Kobayashi Maskawa Institute,
Nagoya University, Nagoya 464-8602, Japan}

\date{\today}


\begin{abstract}

We investigate the motion of a massive particle around a spherically symmetric black hole surrounded by a stationary and radial inflow of perfect fluid. The background spacetime is modelled as a spherically symmetric solution to the Einstein field equations, where the effect of the fluid on the geometry is treated as a perturbation on the Schwarzschild background. The equation of state for the fluid is assumed to follow the linear relationship $p = w \rho$, where $p$ is the pressure, $\rho$ is the energy density with $w$ being a constant. 
The stress-energy tensor is treated as a phenomenological model to capture deviations from the vacuum Einstein theory. We allow the parameter $w$ of the equation of state to take both positive and negative values accepting a broad range of scenarios including exotic ones.
Specifically, we examine the cases $w =2/3$, $1/3$, $-3/4$ and $-4/3$. For $\rho\geq0$, the former two cases satisfy all standard energy conditions while the case of $w=-3/4$ violates the strong energy condition and the case of $w=-4/3$ violates all standard energy conditions. 
By solving the geodesic equations, we visualize the time-like geodesics around the black hole, focusing on the apsis shift of the orbit. To gain further insight into the effects of accretion, we employ the method of osculating orbital elements. Additionally, we analyze the observable effects on spacetime by studying the redshift of the orbiting test particles as an example of possible observables.
We show that the difference in the particle orbits due to the matter accretion may be probed by using the redshift observation of stars orbiting around the black hole.

\end{abstract}

\maketitle  

\section{Introduction} \label{sec:intro}

Black holes, astrophysical objects first predicted theoretically, have always attracted much interest. 
They provide a theoretical playground for investigating the fundamental principles of gravity and spacetime geometry. 
Because the geometry around a black hole is shaped by the surrounding matter and the underlying gravitational theory, studying this geometry allows us to explore not only the black hole's environment but also the gravitational theory governing spacetime in regions with extreme gravity that have not been observed before. 
Ideal observations near a black hole offer a unique opportunity to probe these strong gravity regimes.

Recent advancements in observational technology have provided unprecedented insights into the regions surrounding black holes. For example, the detection of gravitational waves from black hole mergers by the LIGO-Virgo-KAGRA collaboration~\cite{PhysRevLett.116.061102, KAGRA:2013rdx} has revealed highly non-linear gravitational dynamics before and after these mergers. Additionally, the imaging of black hole shadows, such as those of the supermassive black holes in M87~\cite{Akiyama_2019} and in our own Galaxy~\cite{EventHorizonTelescope:2022wkp}, offers valuable information about the immediate environments of these extreme objects. Moreover, long-term observations of stars orbiting the Sgr A* black hole~\cite{GRAVITY:2020gka} have proven to be one of the simplest and most effective methods for observational study of black holes.
If future observations with the Square Kilometer Array (SKA)~\cite{Smits:2008cf} discover pulsars orbiting Sgr A*, these observations could become a powerful tool for testing theoretical predictions related to black hole systems.
Despite these advancements, the nature of environments near black holes remains elusive, and ongoing observational studies aim to further test general relativity in strong gravity regimes~\cite{Will2014}.
%
%

In this study, we consider the effects of matter accretion into a black hole on the orbit of a test particle. 
Given that ordinary matter comprises less than $5\%$ of the universe's total energy density according to Planck data \cite{2020, Semiz_2022,semiz2022}, exploring models of exotic matter—including various forms of dark energy and phantom energy—has become imperative \cite{Novosyadlyj_2013}. 
Therefore we take an open-minded 
approach to the matter field, allowing for the possibility of exotic matter accretion alongside ordinary matter that satisfies the standard energy conditions.
In cosmology, dark energy may exhibit an exotic equation of state~(see, e.g., Ref.~\cite{DESI:2024mwx} for recent observational constraints), although its true nature remains unknown. In such cases, an exotic equation of state could imply modifications to the underlying gravitational theory rather than the presence of an extraordinary matter field. It is possible that a similar situation could arise in observations of the environments surrounding black holes. Therefore, examining the effects of relativistic exotic matter around black holes offers an intriguing avenue for exploration \cite{tan2024}.

In our phenomenological approach to matter accretion, we consider the stress-energy tensor of the accretion process as an effective stress-energy tensor, $T_{\mu\nu}^{\rm eff}$, which also accounts for potential modifications to the gravitational theory. Formally, this can be written as
\begin{equation}
    G_{\mu\nu} + M_{\mu\nu} =  8\pi T_{\mu\nu} 
    \longrightarrow  G_{\mu\nu} = 8 \pi T_{\mu\nu}^{\rm eff}:=  8\pi T_{\mu\nu}-M_{\mu\nu},
    \label{eq:Gmunu}
\end{equation}
where $G_{\mu\nu}$, $M_{\mu\nu}$, and $T_{\mu\nu}$ represent the Einstein tensor, the deviation due to modifications in the gravitational theory, and the stress-energy tensor of ordinary matter, respectively.
It is important to note that although $T^{\rm eff}_{\mu\nu}$ includes the modification term $M_{\mu\nu}$, it must still satisfy the standard equation of motion $\nabla^\mu T^{\rm eff}_{\mu\nu} = 0$, as the left-hand side satisfies the Bianchi identity. One of the simplest ways to probe the effects of $T^{\rm eff}_{\mu\nu}$ is by examining the geodesic motion of test particles, which is the focus of this paper.

For simplicity, we assume spherical symmetry and model the effective stress-energy tensor $T^{\rm eff}_{\mu\nu}$ as a perfect fluid. Following the cosmological convention, we adopt the linear equation of state $p = w \rho$, where $p$ is the pressure, $\rho$ is the energy density, and $w$ is a constant. Specifically, we consider the cases
$w = 2/3$, 
$1/3$, $-3/4$, and $-4/3$.
%
For $\rho\geq0$, the former two cases satisfy all standard energy conditions while the case of $w=-3/4$ violates the strong energy condition and the case of $w=-4/3$ violates all standard energy conditions.
First, we need to construct a solution for the system in which we consider the perfect fluid accretion onto the black hole while accounting for the effects of the accreting fluid on the spacetime geometry. However, we are primarily interested in scenarios where the dominant gravitational influence is the black hole itself, and the effects of $T^{\rm eff}_{\mu\nu}$ are small perturbations.
Therefore, to avoid the complexities of obtaining fully non-linear solutions, we construct the background spacetime perturbatively, assuming the effects of $T^{\rm eff}_{\mu\nu}$ are sufficiently small, following Refs.~\cite{Babichev_2012,Kimura:2021dsa}. This perturbative approach allows us to focus on the modifications introduced by the accreting fluid
in the background Schwarzschild spacetime. 

The investigation of characteristic behaviours in apsis shifts in the presence of matter distributions around black holes is particularly significant~\cite{Nucita_2007}.
There are a few previous works in which exotic matter fields are also considered assuming spherically symmetric static configurations~\cite{Iwata_2016,Harada:2022uae}. 
However, when we describe the surrounding matter field by a perfect fluid, under the existence of a black hole, a static configuration is not necessarily expected. 
In addition, the spherically symmetric assumption may lead to an unexpected singular behaviour that prevents the construction of the global spacetime solution~\cite{Iwata_2016}.
In this paper, we consider a spherically symmetric stationary flow of a perfect fluid which describes the effective stress-energy tensor given in Eq.~\eqref{eq:Gmunu}. 
Under this setup, we can comprehensively investigate the effects of the accretion on a test particle motion
and the impact of different equations of state.

The organization is as follows. 
In the second section, we derive a spacetime metric 
that includes the matter field's backreaction. 
In the third section, we analyze the geodesic dynamics of a test particle in the
spacetime constructed in Sec.\ref{sec:style}.  
The Osculating Orbital Element method 
is introduced in Sec.~\ref{sec:OOE}
to evaluate the change in the apsis 
on the test particle trajectory.
In Sec.~\ref{sec:redshift}, as a demonstration of an observable effect, we calculate the redshift of the test particle 
considering an edge-on observer.
Finally 
Sec.~\ref{sec:sumcon} is devoted to a summary and conclusions. 
This paper uses geometrized units $(c=G=1)$ and a metric signature $(-,+,+,+)$. 

\section{Spherically symmetric spacetime perturbed with a stationary perfect fluid flow} \label{sec:style}
\subsection{Equations of motion for a general spherically symmetric spacetime}

We work with a spherically symmetric line element expressed as
\begin{equation}
    d s^2 = -e^{\nu(V,r) + 2\lambda(V,r)}d V^2 + 2 e^{\lambda(V,r)}d V d r + r^2 d \Omega^2, 
    \label{eq:sphmet}
\end{equation}
where $V$ and $r$ are the ingoing null and radial Eddington-Finkelstein-like coordinates. $d\Omega^2$ represents the metric of a unit 2-sphere. $\nu(V,r)$ and $\lambda(V,r)$ are functions that depend on both $V$ and $r$.
The full set of equations then is given by the Einstein equations
\begin{equation}
    G_{\mu \nu}[g_{\mu \nu}] = 8 \pi T_{\mu \nu}[g_{\mu \nu}],
    \label{eq:EinsteinEquation0}
\end{equation}
and the equations of motion for the surrounding fluid
\begin{equation}
    \nabla^\mu T_{\mu\nu}[g_{\mu \nu}]=0, 
\end{equation}
where the dependence on the metric is explicitly indicated. By substituting the spherically symmetric metric \eqref{eq:sphmet} into the Einstein equations \eqref{eq:EinsteinEquation0}, we obtain a set of differential equations for $\nu(V,r)$ and $\lambda(V,r)$ as follows: 
\begin{eqnarray}
\label{eq:t00}
    8\pi T_{~0}^0 &=& e^{\nu} \left( \frac{1}{r^2} + \frac{\nu'}{r} \right) - \frac{1}{r^2},
\\
\label{eq:t10}
    8\pi T_{~0}^1 &=&  - \frac{e^{\nu}}{r} \dot{\nu},
\\
\label{eq:t01}
    8\pi T_{~1}^0 &=&  - \frac{2 (e^{-\lambda})'}{r}, 
\\
\label{eq:t11}
        8\pi T_{~1}^1 &=& e^{\nu} \left( \frac{1}{r^2} +\frac{\nu'}{r} + \frac{2 \lambda '}{r} \right) - \frac{1}{r^2}, 
\\
\label{eq:t22}
        8\pi T_{~2}^2 &=& 8\pi T_{~3}^3 = e^{\nu} \left(  \lambda'' + \frac{\nu''}{2} + \lambda'^2 +\frac{\nu'^2}{2} + \frac{\lambda'+\nu'}{r} + \frac{3}{2} \lambda' \nu' \right) + e^{-\lambda} \dot{\lambda}' ,
\end{eqnarray}
where the dot notation represents $\partial/ \partial V$ and the prime represents $\partial/ \partial r$. 
The Eqs.~\eqref{eq:t00}, \eqref{eq:t10}, and \eqref{eq:t01} form a set of independent equations, while \eqref{eq:t11} and \eqref{eq:t22} can be derived from the others.

\subsection{Perturbed spacetime}

To obtain a spherically symmetric background spacetime metric under the influence of an accreting matter field, we follow the perturbation method described in \cite{Babichev_2012}. Here we briefly review the procedure.

First, for the zero-order approximation, we consider a vacuum solution without an accreting matter field
    \begin{equation}
        G_{\mu\nu}[g^{(0)}_{\mu\nu}] = 0.
    \label{eq:Sch0}
    \end{equation}
Since we are focusing on spherically symmetric spacetimes, the vacuum solution can be described by the Schwarzschild solution as
\begin{equation}
    d s^2 = -\left(1-\frac{2M_0}{r}\right)d V^2 + 2 d V d r + r^2 d \Omega^2, 
    \label{eq:Sch}
\end{equation}
where $M_0$ is the mass of the Schwarzschild black hole.

For the first-order approximation we consider
\begin{equation}
    \label{eq:EinsteinEquation}
    G_{\mu\nu}[g^{(0)}_{\mu\nu} + \kappa g^{(1)}_{\mu\nu}] = 8\pi T_{\mu\nu} [g^{(0)}_{\mu\nu}],
\end{equation}
and
\begin{equation}
\nabla^\mu T_{\mu\nu}[g^{(0)}_{\mu \nu}]=0,   
\end{equation}
where we have introduced the book-keeping parameter $\kappa$ 
representing the order of the perturbation due to the existence of the matter field. 
The matter density must be sufficiently sparse to be introduced as a perturbation.
For convenience, we replace $\nu(V,r)$ for the function $M(V,r)$ to emulate the zeroth order form as
\begin{equation}
    e^{\nu(V,r)} \equiv 1- \frac{2 M(V,r)}{r}.
\end{equation}
Assuming a spherically symmetric stationary matter distribution, \(T_{~\mu}^{\nu} = T_{~\mu}^{\nu}(r)\), from Eqs.~\eqref{eq:t00} and \eqref{eq:t10} we obtain the expressions for the partial derivatives of $M(V,r)$ in terms of $V$ and $r$ as 
\begin{eqnarray}
\label{eq:mdash}
    M'&=& -4\pi T_{~0}^0 r^2,  
    \\
\label{eq:mdot}
Q:=    \dot{M} &=& 4 \pi T_{~0}^1 r^2, 
\end{eqnarray}
where we have introduced the parameter of the accretion rate, denoted as $Q$.
By taking partial derivatives of Eq.~\eqref{eq:mdash} for time $t$ and Eq.~\eqref{eq:mdot} for $r$, and taking into account the time-independence of $T_{~0}^0$ and $T_{~0}^1$, we can show that $Q=\dot{M}$ is constant. 
We integrate both \eqref{eq:mdash} and \eqref{eq:mdot} to obtain a full expression of \(M(V,r)\) as
\begin{equation}
    M(V,r) = M_0 + Q V - 4\pi \int_{r_0}^r T_{~0}^0(r) r^2 dr,
    \label{eq:massf}
\end{equation}
where \(r_0\) is an arbitrary radius, we have set as the initial radius of particle motion $r_i=120M_0$. We have also introduced \(M_0\) as an integration constant, so that \(M(V,r)\) is equal to the mass of the Schwarzschild background at \(r = r_0 = 120 M_0\).
In a similar manner, by partially integrating Eq.~\eqref{eq:t01} for \(r\) and \(V\), we obtain 
\begin{equation}
    \lambda(V,r) \left(1+\mathcal O(\kappa)\right) =  4\pi \int_{r_0}^{r} T_{~1}^0 r dr + S(V), 
\end{equation}
where \(S(V)\) is a function that depends only on \(V\). We can conveniently set it to zero by redefining the time coordinate \(V\) as \(e^{S(V)}d V \rightarrow d V\).

So far, we have obtained explicit expressions for the two metric functions \(M(V,r)\) and \(\lambda(V,r)\) up to the first order of the matter contribution. For the consistency of our perturbative treatment, we need to impose the following restrictions 
\begin{eqnarray}
    |Q V|\ll&& M_0,
    \label{eq:lim1}
\\    
     \left| 4 \pi \int_{r_0}^{r} T_{~0}^0 (r) r^2 dr \right| \ll&& M_0,
     \label{eq:lim2}
\\
    \left| 4\pi \int_{r_0}^r T_{~1}^0 r dr \right| \ll&& 1.
    \label{eq:lim3}
\end{eqnarray}

The first expression \eqref{eq:lim1} states that the value of \(|QV|\)—the total mass accreted over a given period (from 0 to V)—must be significantly smaller than \(M_0\). In other words, the mass accreted will never approach the mass of the black hole during the timeframe we are examining. 
The condition (\ref{eq:lim2}) indicates that the contribution of the time-independent density distribution of the matter field to the total mass inside the radius $r$ must be much smaller than the background black hole mass $M_0$. 
The last restriction (\ref{eq:lim3}), can be rephrased as $|\lambda|\ll 1$, keeping our correction to the Schwarzschild metric sufficiently small. 
Finally, all of these conditions enable us to simplify the metric as
\begin{equation}
\label{eq:PerturbedMetric}
    ds^2 \simeq -\left(1-\frac{2 M(V,r)}{r}\right)(1+2\lambda(r))dV^2 + 2(1+\lambda(r))dV dr + r^2 d\Omega^2,
\end{equation}
up through the first order of $\kappa$. 

\subsection{Conservation law for a perfect fluid}

Let's consider a case of a perfect fluid with a stress-energy tensor
\begin{equation}
    T_{\mu\nu} = (\rho + p) u_{\mu} u_{\nu} + p g_{\mu\nu}, 
\end{equation}
where $p$, $\rho$ and $u^\mu$ are the pressure, energy density and fluid four-velocity, respectively.
We consider a radial ingoing fluid, then by using the normalization condition for the Schwarzschild metric, with a spherically symmetric assumption, the four-velocity can be written as 
\begin{equation}
    u^{\mu} = \left(\frac{1}{\sqrt{f_0 + v^2}+v}, -v, 0, 0\right),
\end{equation}
where $v$ corresponds to
the radial infalling velocity of the fluid, $r$ is the radial coordinate and $f_0 = 1 - 2 M_0/r$.
We note that the radial fluid velocity $\bar v$ measured by the Killing observer, whose four-velocity is proportional to $(\partial/\partial V)^\mu$ is given by\footnote{
The Killing observer four-velocity $\bar u^\mu$ is given by 
$$\bar u^\mu=\frac{1}{\sqrt{f_0}}\left(\frac{\partial}{\partial V}\right)^\mu,$$
and the radial unit form $\left(e^r\right)_\mu$ normal to $\bar u^\mu$ is given by 
$$\left(e^r\right)_\mu=\frac{1}{\sqrt{f_0}}\left(dr\right)_\mu. $$
Then we define $\bar v$ as
$$\bar v=\frac{u^\mu\left(e^r\right)_\mu}{u^\mu\bar u_\mu}. $$
}
\begin{equation}
    \bar v=\frac{v}{\sqrt{f_0+v^2}}. 
\end{equation}

In this study, we consider an inflowing fluid and positive energy density, meaning $v>0$ and $\rho>0$. 
The elements of the energy-momentum tensor are
\begin{eqnarray}
    T^0_{~0} &=&\frac{-\rho \sqrt{f_0 + v^2} + p v}{\sqrt{f_0 + v^2}+v} ,
 \\
 T^0_{~1} &=& \frac{\rho + p}{(\sqrt{f_0 + v^2}+v)^2} ,
 \\
 T^1_{~0}& =& (\rho + p) v \sqrt{f_0 + v^2} ,
 \\
 T^1_{~1} &=&\frac{\sqrt{f_0 + v^2} p - v \rho}{\sqrt{f_0 + v^2}+v},
 \\
 T^2_{~2} &=& T^3_3 = p.
\end{eqnarray}
Considering the conservation of energy-momentum $\nabla_\nu T^{\nu}_{~\mu} = 0$, 
we can get a set of equations for
$\mu = 0$ and $\mu = 1$ as follows:
\begin{eqnarray}
\label{eq:mu0}
&&\frac{\partial}{\partial r} \ln{(T^1_{~0} r^2)} = 0 ,
\\
\label{eq:mu1}
&&\frac{\partial}{\partial r} [\rho (r^2 v)^{1+\omega}] = 0 .
\end{eqnarray}
From Eq.~\eqref{eq:mu0}, we find that $T^1_{~0} r^2$ is constant. This is consistent with the definition of $Q$ given in Eq.~\eqref{eq:mdot}. As a result, we can express $T_{~0}^1$ in terms of $Q$ and $r$ from Eq.~\eqref{eq:mdot}. We can integrate and rephrase Eqs.~\eqref{eq:mu1} and \eqref{eq:mu0} as
\begin{align}
    \label{eq:A}
    4 \pi r^2 (1 + w) \rho v \sqrt{f_0 + v^2} &= Q, \\
    \label{eq:B}
    \rho (r^2 v)^{1 + w} &= B, 
\end{align}
where $B$ is an integration constant. 
It is important to note that, due to the positive nature of $\rho$ (and by Eq.~\eqref{eq:B} also of B), the sign of the accretion rate $Q$ is determined by the linear term $(1+w)$ in Eq.~\eqref{eq:A}.

For a given value of $w$, we can eliminate $\rho$ using Eqs.~(\ref{eq:A}) and (\ref{eq:B}) to derive an implicit functional form of $v$ as a function of $r$: 
\begin{equation}
    \label{eq:QB}
    F(r,v):= \\4 \pi (1 + w)(r^2 v)^{- w} \sqrt{f_0 + v^2} =  \frac{Q}{B}.
\end{equation}
\begin{figure}[ht!]
    \centering
    \subfloat[Case 1: \(w=1/3\)]{
        \includegraphics[width=0.48\linewidth]{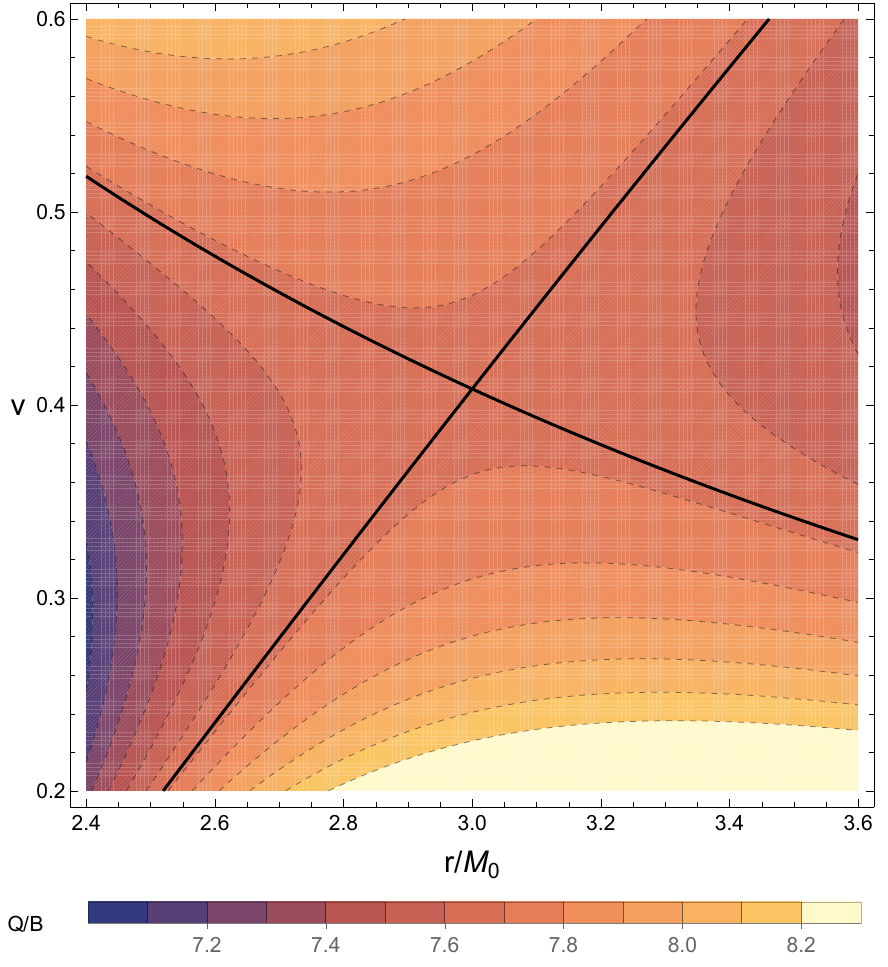}\label{im:cp1}}
    \hfill
    \subfloat[Case 2: \(w=2/3\)]{
        \includegraphics[width=0.48\linewidth]{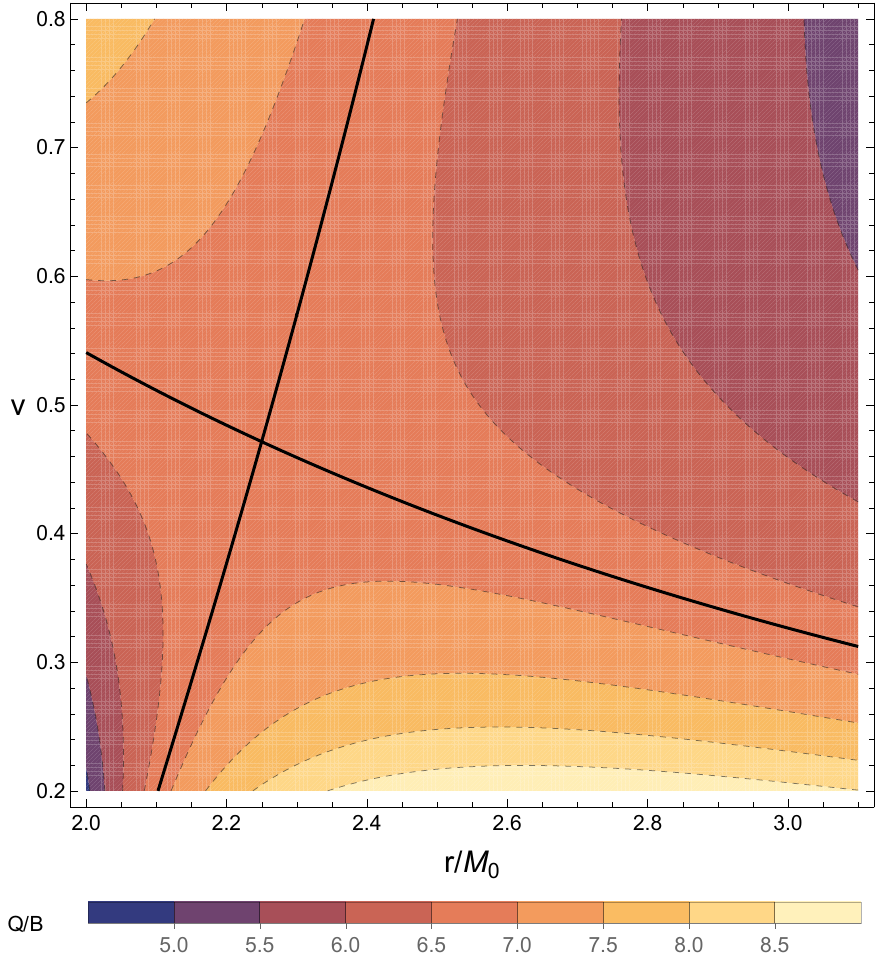}
        \label{im:cp2}
        }
    \vspace{0.5cm} 
    \subfloat[Case 3: \(w=-3/4\)]{
        \includegraphics[width=0.48\linewidth]{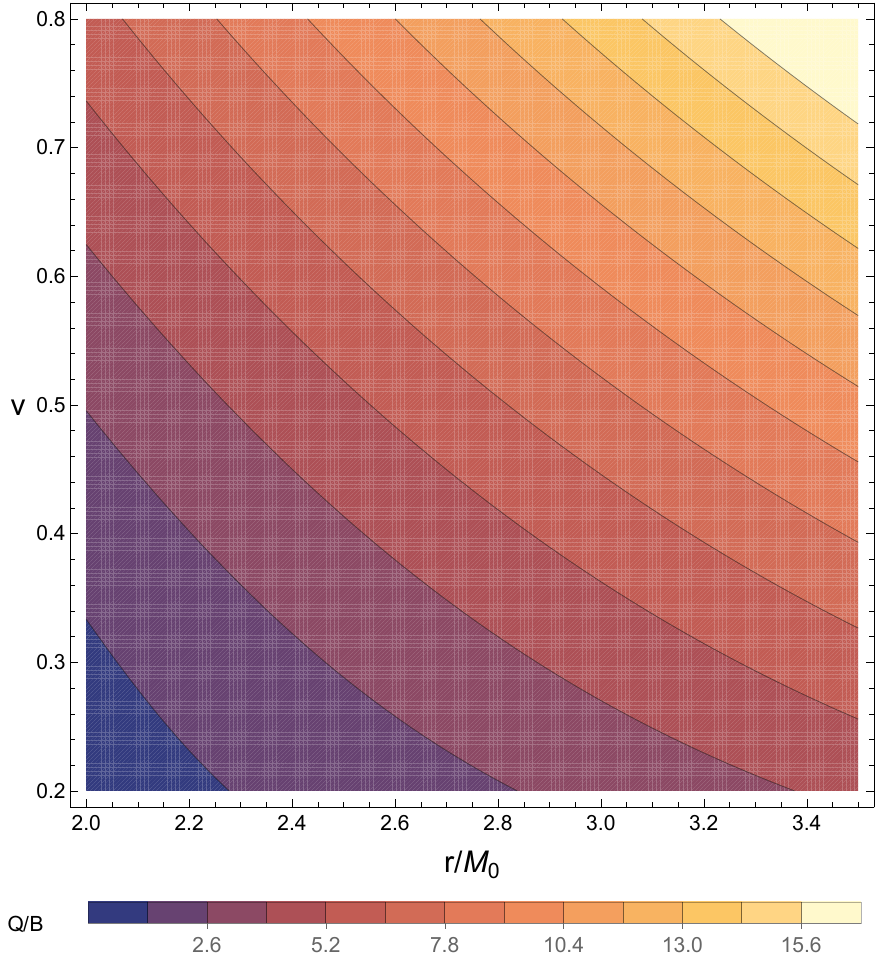}
        \label{im:cp3}
        }
    \hfill
    \subfloat[Case 4: \(w=-4/3\)]{
        \includegraphics[width=0.48\linewidth]{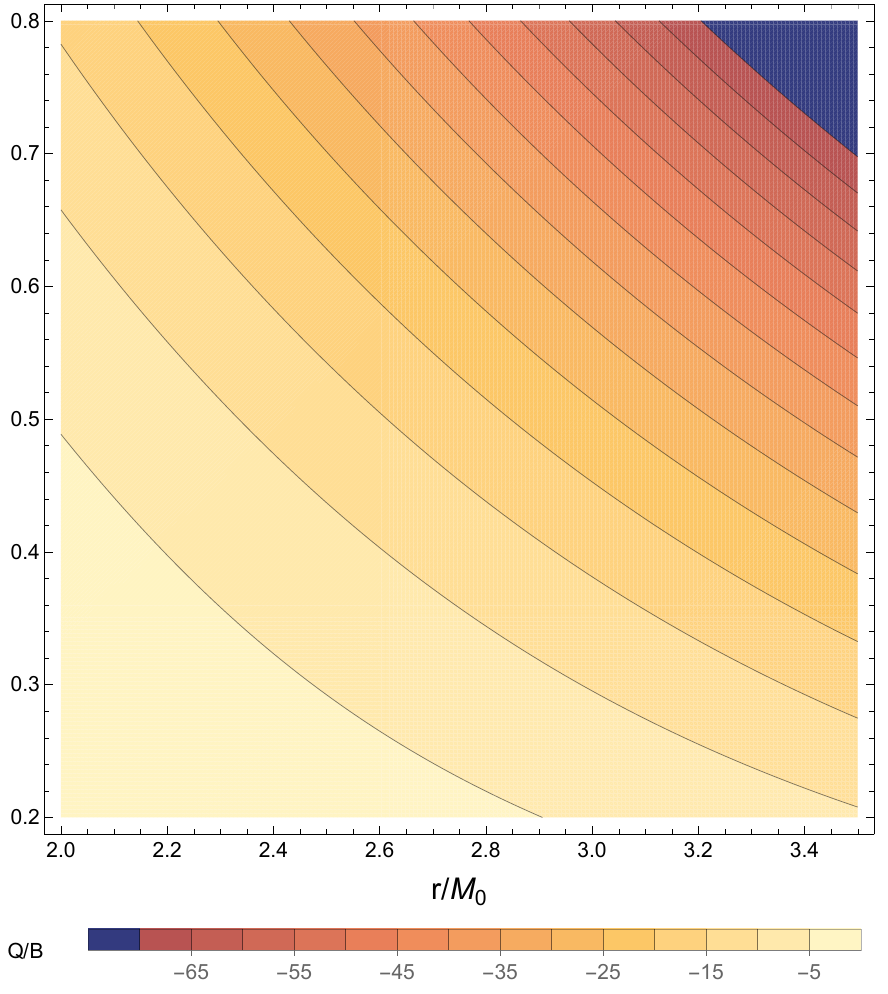}
        \label{im:cp4}
        }
    \caption{Contour plot of Eq.~\eqref{eq:QB} for different right-hand side constants.}
    \label{fig:group1}
\end{figure}
In Figure~\ref{fig:group1}, we present the contour map of the ratio $Q/B$ as a function of $r$ and $v$ for the values $w = \frac{1}{3}$,  $\frac{2}{3}$,
$-\frac{3}{4}$, and $-\frac{4}{3}$ respectively. For cases where $w > 0$, we identify a saddle point representing the critical value of $F(r, v)$. At this critical point, the fluid velocity $\bar v$ measured by the Killing observer corresponds to the sound speed $v_s=\sqrt{\frac{dp}{d\rho}}=\sqrt{w}$, as referenced in \cite{1972Curtis}. 
From the condition $\partial_r F|_{(r,v)=(r_{cr},v_{cr})}=\partial_v F|_{(r,v)=(r_{cr},v_{cr})}=0$, we obtain 
\begin{eqnarray}
    r_{cr}&=&\frac{1+3w}{2w}M_0,
    \\
    v_{cr}&=&\sqrt{\frac{w}{3w+1}}.
\end{eqnarray}
Therefore, for $w>0$, we obtain
\begin{equation}
    (Q/B)_{cr} =F(r_{cr},v_{cr})= 4^{1 + w}\pi M_0^{-2w} (1+w)(1+3 w)^{-1/2}\left( \frac{w}{1+3w}\right)^{3w/2}.
\label{eq:QBcritical}
\end{equation}
In the case $w=1/3$ (Fig~\ref{im:cp1}), the critical value of the velocity $\bar v$ is given by $\sqrt{w}=1/\sqrt{3}$ at $r_{cr}= 3M_0$~\cite{Koga:2016jjq,Koga:2018ybs,Koga:2019teu,Tsuchiya:2020suk}.
For the transonic accretion flow with $w>0$, the value of $v$ increases with decreasing $r$. For $w < 0$, no saddle point exists for $F(r,v)$ 
as observed in Fig.~\ref{im:cp3} and \ref{im:cp4}, 
and the qualitative behaviours of $v$ are similar to the critical solutions for positive $w$ cases for all contours in the $r$-$v$ plot.

\section{General Formulation of the Geodesic Equation} \label{sec:floats}

Let's analyze the geodesic motion of test particles in the perturbed metric. Due to the spherical symmetry of the system, we can simplify our analysis by restricting the trajectory to the plane where $\theta = \frac{\pi}{2}$. In this context, the Lagrangian describing the motion of the test particle is given by 
\begin{equation}
    L =\frac{1}{2}g_{\mu\nu}\frac{dx^\mu}{d\tau}\frac{dx^\nu}{d\tau} =\frac{1}{2} \left[-\left(1 - \frac{2 M(V,r)}{r}\right) (1 + 2 \lambda) \left( \frac{dV}{d\tau}\right)^2 +2(1+\lambda)\left( \frac{dV}{d\tau}\right)\left( \frac{dr}{d\tau}\right) + r^2 \left( \frac{d\phi}{d\tau}\right)^2 \right], 
\end{equation}
where $g_{\mu\nu}$ is a metric component and $\tau$ is the 
proper time. 
Since $\phi$ is a cyclic coordinate, we can introduce the angular momentum $l$ as a constant of motion as
\begin{equation}
   l = r^2  \frac{d\phi}{d\tau}. 
    \label{eq:const}
\end{equation}
From the Euler-Lagrange equation for the radial coordinate $r$, we obtain
\begin{equation}
\label{eq:radius}
    (1+\lambda) \frac{d^2 V}{d \tau^2} + \left( 
 \frac{M}{r^2} +4 \pi T_{~0}^0 r + 4 \pi T_{~1}^0 r - 8\pi M_0 T_{~1}^0  
 + \frac{2 M_0\lambda}{r^2} \right) \left( \frac{dV}{d\tau} \right)^2 
 - \frac{l^2}{r^3} 
 \simeq
 0, 
\end{equation}
where we have dropped the higher order terms concerning  $\kappa$.  Similarly, for the time coordinate $V$, we obtain
\begin{multline}  
\label{eq:time}
    \frac{Q}{r}(1 + 2\lambda) \left( \frac{dV}{d\tau} \right)^2 + \left( -\frac{2 M}{r^2} - \frac{4 M_0 \lambda}{r^2} - 8\pi T_{~0}^0 r - 8\pi T_{~1}^0 r + 16 M_0 \pi T_{~1}^0 \right) \left( \frac{dr}{d\tau} \right) \left(\frac{dV}{d\tau} \right) \\ - \left( 1 - \frac{2M}{r}+ 2\lambda  - \frac{4M_0\lambda}{r} \right) \frac{d^2V}{d \tau^2} + 4\pi T_{~1}^0 r \left( \frac{d r}{d \tau} \right)^2 + (1 + \lambda)\frac{d^2r}{d\tau^2} 
 \simeq
    0. 
\end{multline}

By using the normalization condition 
$g_{\mu \nu} (dx^{\mu}/d\tau) (dx^{\nu}/d\tau)  = -1$
and Eq.~\eqref{eq:const}, 
we obtain an additional equation
\begin{equation}
\label{eq:metric}
    \left( 
 1 - \frac{2M}{r} + 2\lambda -  \frac{4 M_0\lambda}{r} \right) \left( \frac{dV}{d\tau} \right)^2 - 2(1+\lambda) \left(\frac{d V}{d \tau}\right) \left(\frac{d r}{d \tau}\right) 
 - \frac{l^2}{r^2}-1 
 \simeq
 0.
\end{equation}

Choosing the angle coordinate $\phi$ as the independent variable is more convenient for our purposes. Using \eqref{eq:const} and substituting the radial coordinate $r$ with $u:=1/r$, we obtain

\begin{align}
    0&\simeq-\left(1 - 2 M u + 2\lambda - 4 M_0 \lambda u\right)\left(\frac{d V}{d \phi}\right)^2 
    -\frac{2(1+\lambda)}{u^2}\left(\frac{d V}{d \phi}\right)\left(\frac{d u}{d \phi}\right) + u^{-2} + l^{-2} u^{-4}, 
    \label{eq:firstgeodesic} \\
    \frac{d^2 u}{d \phi^2} &\simeq 
    -(1 - 2 M u +\lambda - 2 
    M_0 \lambda u) u^2 \frac{d^2 V}{d \phi^2} + Q u^3 \left( \frac{d V}{d\phi}\right)^2 + \frac{4 \pi T_{~1}^0}{u^3} \left(\frac{d u}{d \phi}\right)^2
    \nonumber \\
    & \quad  - \left[2 u - 6 M u^2 + 2 \lambda u - 6 M_0 \lambda u^2 - \frac{8 \pi (T_{~0}^0 + T_{~1}^0)}{u} + 16 \pi T_{~1}^0 M_0 \right] \left( \frac{d u}{d \phi}\right) \left( \frac{d V}{d \phi}\right),
    \label{eq:secondgeodesic} \\
    \frac{d^2 V}{d \phi^2} &\simeq 
    \Bigg(- M u^2 - \frac{4 \pi (T_{~0}^0+ T_{~1}^0)}{u} 
    + 8 \pi T_{~1}^0 M_0 -  M_0 \lambda u^2\Bigg)\Bigg( \frac{dV}{d\phi} \Bigg)^2 
    - \frac{2}{u} \left(\frac{d V}{d \phi}\right) \left(\frac{d u}{d \phi}\right) + \frac{1-\lambda}{u}. 
    \label{eq:thirdgeodesic}
\end{align}

\begin{figure}[htbp]
    \centering
    \includegraphics[width=0.48\linewidth]{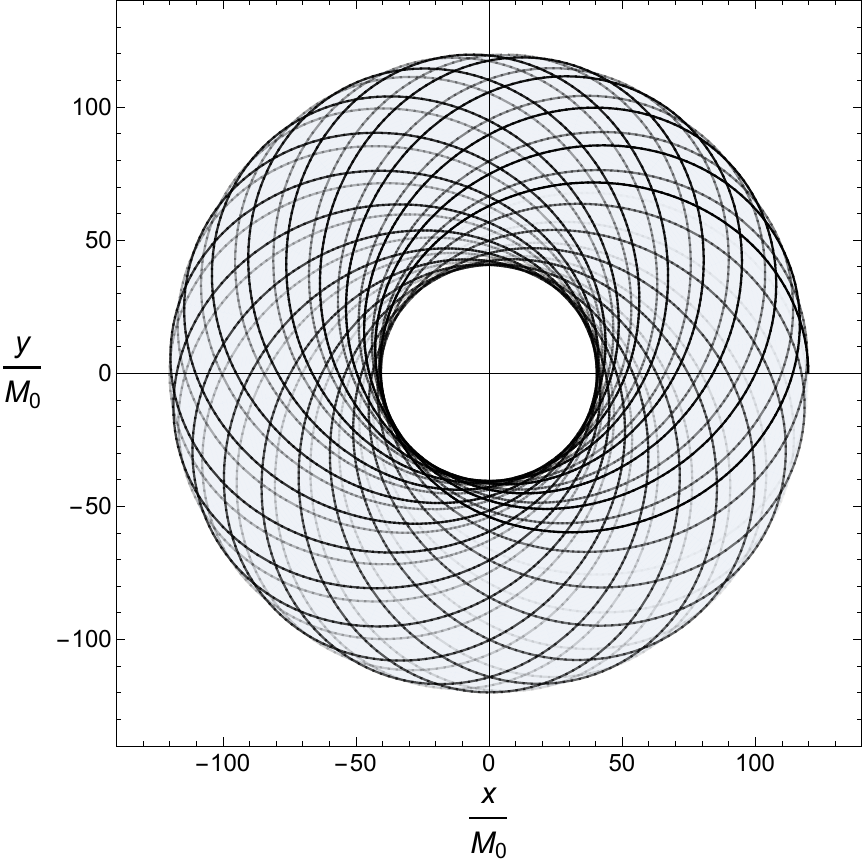}
    \caption{Polar plot showing the vacuum geodesic solution for a test particle orbiting a black hole, highlighting the orbital trajectory in a vacuum environment.}
    \label{fig:polarplot0}
\end{figure}

By taking $Q = 0$ and $T^\mu_{~\nu}=0$, and therefore $M=M_0$ and $\lambda=0$, we can reduce the previous set of equations (\ref{eq:firstgeodesic}, \ref{eq:secondgeodesic}, \ref{eq:thirdgeodesic}) back to the geodesic equations for the Schwarzschild spacetime as 
\begin{equation}
    \frac{d^2 u}{d \phi^2} - \frac{M_0}{l^2}  + u - 3 M_0 u^2 = 0 .
    \label{eq:Sc0}
\end{equation}
To illustrate the particle motion described by Eq.~\eqref{eq:Sc0} with the parameters $l=8$ and $M_0=1$, we present a polar plot in Figure \ref{fig:polarplot0}. As is well known, the trajectory does not resemble a closed Keplerian orbit, and we can observe the precession of the apsis due to relativistic effects. This reference will serve as a baseline for comparing the perturbed geodesic motion discussed in the subsequent sections of this paper.

To begin, we will numerically solve the geodesic equations to understand how fluid accretion affects geodesic motion based on the parameter \( w \). We will use the following parameters for the test particle's movement: angular momentum \( l = 8 M_0 \) and an initial radius of \( r = r_i = 120 M_0 \). 
\begin{figure}[ht!]
    \centering
    \subfloat[Case 1: \(w=1/3\)]{
        \includegraphics[width=0.48\linewidth]{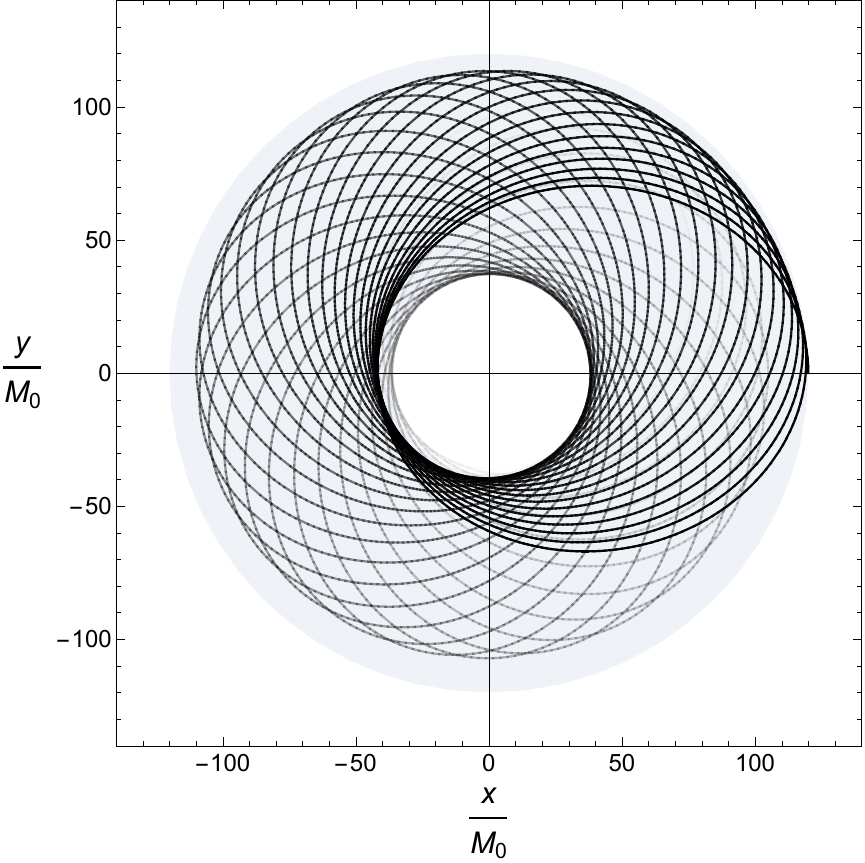}
        \label{fig:polar13}
        }
    \hfill
    \subfloat[Case 2: \(w=2/3\)]{
        \includegraphics[width=0.48\linewidth]{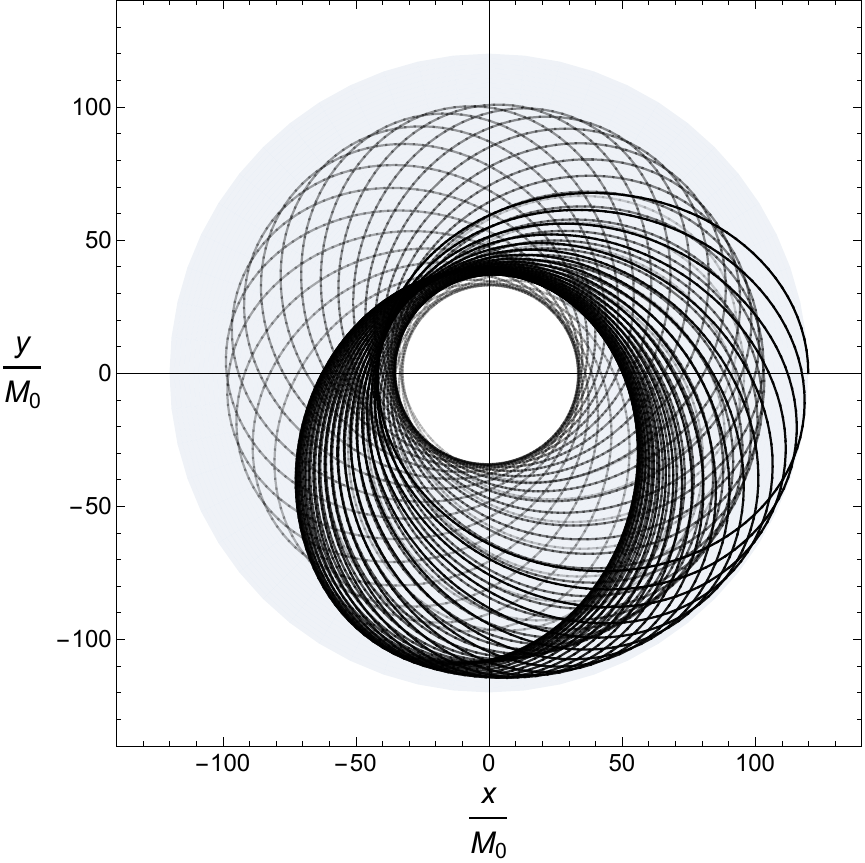}
        \label{fig:polar23}
        }
    \vspace{0.5cm} 
    \subfloat[Case 3: \(w=-3/4\)]{
        \includegraphics[width=0.48\linewidth]{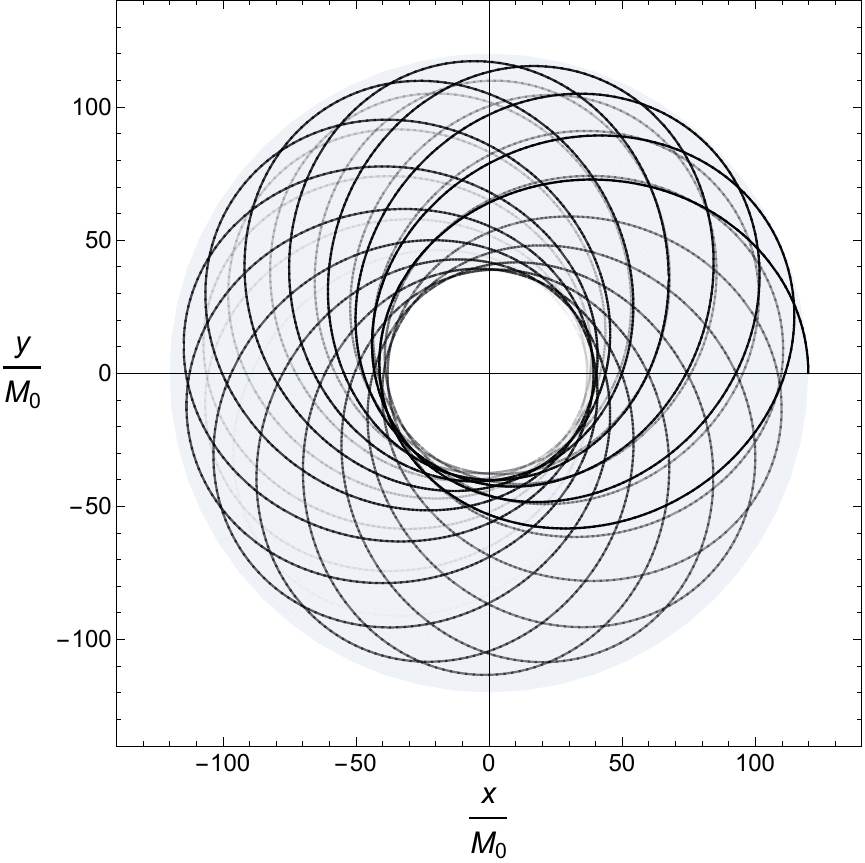}
        \label{fig:polar34}
        }
    \hfill
    \subfloat[Case 4: \(w=-4/3\)]{
        \includegraphics[width=0.48\linewidth]{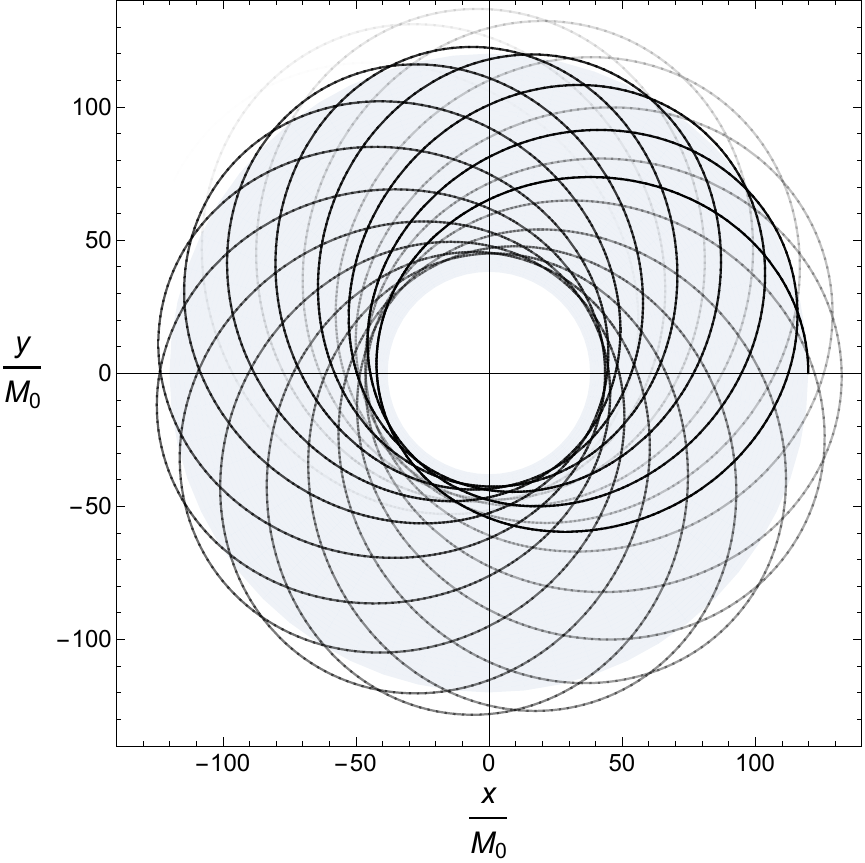}
        \label{fig:polar43}
        }
    \caption{Geodesic orbits \(r(\varphi)\) in polar coordinates for different cases of $w$.}
    \label{fig:groupA}
\end{figure}
In cases 1 ($w=1/3$, Fig.~\ref{fig:polar13}) and 2 ($w=2/3$, Fig.~\ref{fig:polar23}), the accreting matter is characterized by a positive equation of state parameter $w>0$. 
As previously stated, to pick the physically reasonable configuration, we need to fix the value of $Q/B$ to the critical value defined by Eq.~\eqref{eq:QBcritical}.
Thus, we need only to set one parameter, $Q$, to specify the background spacetime.
We have set $Q=10^{-6}$ for both cases. 
As time $V$ increases, the mass function $M(V,r)$ grows, indicating that the amount of matter within a fixed radius increases. 
For cases 1 and 2, due to the positive mass accretion, the gravitational potential gradually deepens, and the radius of the particle motion decreases as time passes. In case 2 we can see a more noticeable effect of the accreting matter than in case 1, 
but it initially displays a different periapsis shift direction on its orbit~(see Fig.~\ref{fig:polar23}). 
For cases 3~($w=-3/4$, , Fig.~\ref{fig:polar34}) and 4~($w=-4/3$, Fig.~\ref{fig:polar43}), there is no critical solution. 
Therefore we need to fix the value of $B$ in addition to the value of $Q$ to determine the fluid distribution. We set $B=4\times10^{-10}$ for both cases. 
For case 3, we use $Q=10^{-6}$, and for case 4, we set $Q=-10^{-6}$. 
It is important to note that the sign of $Q$ is given by the sign of $1+w$ as discussed after Eq.~\eqref{eq:B}.
For case 3~($w = -3/4$), the behaviour of the geodesic is similar to that for case 1. 
That is, the orbital radius decreases with time~(see Fig.~\ref{fig:polar34}). 
This is because the mass function $M(V,r)$ increases with both $V$ and $r$ coordinates, similar to cases where $w>0$. 
Case 4 has a clearly different behaviour from its previous counterparts. The amplitude of the orbital radius increases.
In case 4~($w=-4/3$), since $w<-1$, the accretion rate becomes negative, leading to a mass decrease within a given radius over time. Consequently, the geodesic orbital radius increases, as illustrated in Fig.~\ref{fig:polar43}.

\section{Method of osculating orbital elements (OOE)}
\label{sec:OOE}

To gain a deeper understanding, let us analyze the system using the OOE method. This analytical technique allows us to calculate the changes in the orbital elements, specifically the apsis shift, after one orbital cycle. While it is beyond the scope of this paper to detail this method, further explanations can be found in Poisson's textbook  \cite{Poisson_Will_2014}.
The core idea is to approximate the real orbit to an osculating ellipse (Keplerian orbit) that instantaneously matches the actual trajectory. We are interested in the secular variation of the $\omega$ orbital element which represents the apsis displacement during the orbit evolution. We consider that the test point mass around the black hole is moving very slowly ($v_s$) in comparison to the speed of light ($t \sim \tau$), so that
\begin{equation}
    \frac{M_0}{r} \sim \frac{l^2}{r^2} \sim v_s^2 << 1.
\end{equation}

\subsection{Apsis shift in the Schwarzschild solution}

Let's consider the differential equation~\eqref{eq:Sc0}. Taking $\tau$, which is related to $\phi$ through Eq.~\eqref{eq:const}, as a time coordinate we obtain
\begin{equation}
\label{scharchildOOM}
    \frac{d^2r}{dt^2} - \frac{l^2}{r^3} = - \frac{M_0}{r^2} - \frac{3 M_0 l^2}{r^4}.
\end{equation}
Comparing it with the Newtonian case, the last term can be interpreted as a relativistic correction for the radial gravitational force. In Ref.~\cite{Poisson_Will_2014} we can find in equation (3.69e) an expression for the osculating equation by using the inverse square Newtonian force as
\begin{equation}
\frac{d\omega}{d\varphi} \simeq \frac{a^2 (1-e^2)^2}{e M_0} \left( \frac{- \cos{\varphi}}{(1 + e \cos{\varphi})^2}\right) \mathcal R, 
\label{eq:domega0}
\end{equation}
where $\mathcal R$ refers to a purely radial perturbative force inserted into the Newtonian inverse square term. Substituting the correction term of Eq.~\eqref{scharchildOOM} into $\mathcal R$ in Eq.~\eqref{eq:domega0}, we obtain the following osculating equation for the angular displacement of the apsis $\omega$
\begin{equation}
\frac{d\omega}{d\varphi} \simeq \frac{3a^2l^2 (1-e^2)^2}{e r^4} \left( \frac{ \cos{\varphi}}{(1 + e \cos{\varphi})^2}\right),
\label{eq:domega1}
\end{equation}
where $\varphi$ and $e$ are the true anomaly ($\varphi=\phi-\pi$) and eccentricity, respectively. We can express the radius for a Kepler orbit by using the angular momentum $l$, eccentricity $e$, true anomaly $\varphi$, and central mass $M_0$ as
\begin{equation}
r=\frac{l^2}{M_{0} (1+e \cos{\varphi})},    
\label{eq:rvphi}
\end{equation}
making the right-hand side of Eq.~\eqref{eq:domega1} only dependent on the true anomaly $\varphi$.
To calculate the apsis displacement after one orbital cycle, we integrate $\varphi$ from $0$ to $2\pi$ as
\begin{equation}
\Delta\omega_0 = \frac{6\pi l^2}{a^2 (1-e^2)^2}= \frac{6\pi M_0^2}{l^2},
\label{eq:omega0}
\end{equation}
where we have used the Keplerian relation 
$aM_0(1-e^2)=l^2$.

\subsection{Apsis shift in the accreting black hole}

We adopt a similar approach for the perturbed black hole spacetime solution, considering the effects of accretion. First, we must specify the order of the perturbation terms. To assess the impact of accretion, we assume specific orders of magnitude relative to the velocity of the test particle ($v_s$)
\begin{equation}
\dot{M}(V,r),M'(V,r),\lambda(r) \sim v_s^4.
\end{equation}
Then, taking there terms up to $v_s$ from equations \eqref{eq:firstgeodesic}, \eqref{eq:secondgeodesic} and \eqref{eq:thirdgeodesic}, we obtain
\begin{eqnarray}
    \frac{d^2r}{d\tau^2} - \frac{l^2}{r^3} &\simeq& - \frac{M}{r^2} - \frac{3 M_0 l^2}{r^4}  - 4\pi T_{~0}^0 r - 4\pi T_{~1}^0 r - \frac{Q}{r}\cr
    &=&
    - \frac{M_0}{r^2}  - \frac{Q V}{r^2} + \frac{4 \pi}{r^2}\int T_{~0}^0 r^2 dr - \frac{3 M_0 l^2}{r^4}- 4\pi T_{~0}^0 r - 4\pi T_{~1}^0 r - \frac{Q}{r}. 
    \label{eq:d2rdtau}
\end{eqnarray}
It is important to note that in this expression, \( M_0 \) represents the mass 
inside the radius $r=r_0$
at the initial time when \( V = 0 \). However, if we are examining 
$n$-th cycle of the orbit, 
we can substitute \( M_0 \) with 
$M_n$. We will discuss how to define $M_n$ soon.  
Additionally, we should acknowledge that accretion affects the orbital parameters \( e \) and \(a \). 
Thus, we redefine these parameters 
for each cycle as
$e_n$ and $a_n$ instead of the initial values of $e$ and $a$. 

By substituting the expression \eqref{eq:rvphi} into $r$ in the right-hand side of Eq.~\eqref{eq:d2rdtau}, we obtain an equation that only depends on the angle parameter $\varphi$ except for the second term, where we face the dependence on the time parameter $V$. To integrate it, we consider $V$ as the standard time in Kepler motion, we use the following Keplerian relations

\begin{eqnarray}
\cos\varphi&=&\frac{\cos\chi-e}{1-e\cos\chi}, \hspace{0.5cm}
\sin\varphi=\frac{\sqrt{1-e^2}\sin\chi}{1-e\cos\chi}, \hspace{0.5cm}
\frac{d\varphi}{d\chi}=\frac{\sqrt{1-e^2}}{1-e\cos\chi}, \\
    V-T&=&\frac{P}{2\pi}\left(\chi-e\sin\chi\right)=\sqrt{\frac{a^3}{M_0}}\left(\chi-e\sin\chi\right),
\end{eqnarray}
where $\chi$, $T$ and $P$ are the eccentric anomaly, time of the apsis passage and the period of the Kepler elliptic orbit. 
Taking $T=0$, the second term can be rewritten as a function of $\chi$. Then, substituting the given expression into $\mathcal R$ in Eq.~\eqref{eq:domega0}, we can calculate the contribution to the apsis shift as 
\begin{equation}
    \int^{2\pi}_0 d\varphi \left( \frac{d\omega}{d\varphi} \right) _{V} =
        -\frac{a^{3/2} Q}{e M_0^{3/2}}         \int^{2\pi}_0 d\varphi 
\frac{(e-\cos{\chi})(\chi - e \sin{\chi})}{(1 - e \cos{\chi})} =-
        \frac{a^{3/2} Q\sqrt{1-e^2}}{e M_0^{3/2}}         \int^{2\pi}_0 d\chi 
\frac{(e-\cos{\chi})(\chi - e \sin{\chi})}{(1 - e \cos{\chi})^2}=0. 
\end{equation}

After performing similar calculations for other terms, we obtain the apsis shit during one orbital period as

\begin{equation}
\label{OEM2}
 \Delta\omega=  \Delta\omega_0+ \Delta\omega_Q+\Delta\omega_{\rho},
\end{equation}
where
\begin{eqnarray}
\label{eq:omegaQ}
    \Delta\omega_Q &=& \frac{2\pi a (1-e^2) Q}{e^2 M_0}
    \left(1-\frac{1}{\sqrt{1-e^2}} \right), 
\\
\label{eq:omegarho}
    \Delta\omega_{\rho} &=&  -\frac{4\pi}{e M_0}
    \int_0^{2\pi}\cos{\varphi}
    \left( \int^{r(\phi)}_{r_0} T_{~0}^0(r)r^2 \,dr \right) d\varphi
    \nonumber \\
    &+& \frac{4\pi (a(1-e^2))^3}{e M_0}\int_0^{2\pi}\frac{\cos{\varphi}}{(1+e\cos{\varphi})^3} T_{~0}^0(r({\varphi}))  d\varphi
    \nonumber \\
    &+& \frac{4\pi (a(1-e^2))^3}{e M_0}\int_0^{2\pi}\frac{\cos{\varphi}}{(1+e\cos{\varphi})^3} 
    T_{~1}^0
    (r({\varphi}))  d\varphi.
\end{eqnarray}
Eqs.~ \eqref{eq:omega0}, \eqref{eq:omegaQ} and \eqref{eq:omegarho} refer to the contributions of General Relativity, accretion rate, and mass of the accretion fluid respectively. The general relativity shift $\Delta \omega_{0}$ component due to the spacetime curvature always makes a positive contribution, whereas, $\Delta \omega_{Q}$ contributions depend on the sign of $Q$, which can only be negative in the scenario where $w<-1$, because of expression \eqref{eq:QB}. The last component, $\Delta \omega_{\rho}$, arises from the energy density and pressure contributions of the accreting fluid. Its contribution has a more complicated relation with the characterization of the fluid. 
Nevertheless, considering the static ($v\ll1$), uniform ($\rho\sim {\rm const.}$)  and weak field ($f_0\sim1$) limit, since we have $T^0_{~0}\sim-\rho$ and $T^0_{~1}\sim(1+w)\rho$, 
we can obtain 
\begin{equation}
    \Delta \omega_\rho\sim
    -4\pi^2\sqrt{1-e^2}\left(1+3w\right)\frac{a^3}{M_0}\rho. 
    \label{eq:delomrhoappr}
\end{equation}
This expression 
shows that, in this limit, the active gravitational mass density~\cite{Schutz:1985jx} determines the sign of $\Delta\omega_\rho$, which 
is consistent with the previous works~\cite{Iwata_2016,Harada:2022uae}. 
The initial values of $\Delta \omega_0$, $\Delta \omega_Q$ and $\Delta \omega_\rho$ by using the OOE method 
are summarized in Table~\ref{table:omegas}.  
\begin{table}[htbp]
 \centering
\begin{tabularx}{150mm}{C|CCCC}
\hline
$w$ & $1/3$ (Case 1) & $2/3$ (Case 2) & $-3/4$ (Case 3) & $-4/3$ (Case 4)
\\ 
\hline
$Q$ & $10^{-6}$& $10^{-6}$ & $10^{-6}$ & $-10^{-6}$ 
\\
\hline
$B$ & $1.30\times 10^{-7}M_0^{2/3}$& $1.48\times10^{-7}M_0^{4/3}$ & $4\times 10^{-10}M_0^{-3/2}$ & $4\times 10^{-10}M_0^{-8/3}$
\\
\hhline{=|====}
$\Delta\omega_0$ & $~~2.95 \times 10^{-1}$ &  $~~2.95 \times 10^{-1}$ &  $~~2.95 \times 10^{-1}$ & $~~2.95 \times 10^{-1}$ \\\hline
$\Delta\omega_Q$ & $-2.41 \times 10^{-4}$ &  $-2.41 \times 10^{-4}$ &  $-2.41 \times 10^{-4}$ & $~~2.41 \times 10^{-4}$ \\ \hline
$\Delta\omega_\rho$ &  $-2.36 \times 10^{-1}$ &  $-5.29 \times 10^{-1}$ &  $~~7.50 \times 10^{-2}$ & $~~1.14 \times 10^{-1}$ \\  \hline
Total & $~~5.83 \times 10^{-2}$ &  $-2.35 \times 10^{-1} $ &  $~~2.96\times 10^{-1}$ & $4.09\times 10^{-1}$ \\ \hline
\end{tabularx}
\caption{Initial values of $\Delta \omega_0$, $\Delta \omega_Q$ and $\Delta \omega_\rho$ evaluated by using the OOE method. 
These values are calculated using the initial Keplerian orbital parameters $M_0$, $e$ and $a$, which can be evaluated from the initial conditions for the particle motion. }
\label{table:omegas}
\end{table}

\begin{figure}[ht!]
    \centering
    \subfloat[Case 1: \(w=1/3\)]{
        \includegraphics[width=0.48\linewidth]{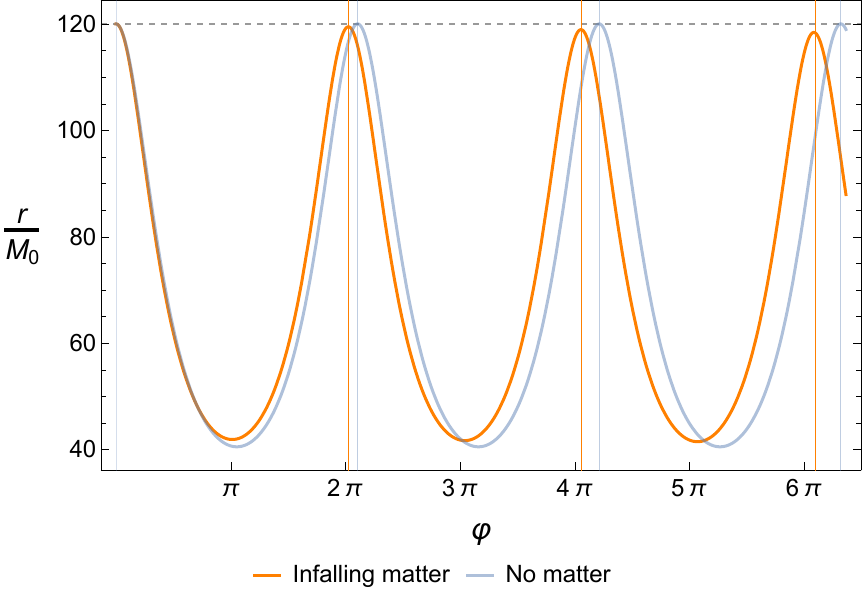}
        \label{fig:shiftC1}
        }
    \hfill
    \subfloat[Case 2: \(w=2/3\)]{
        \includegraphics[width=0.48\linewidth]{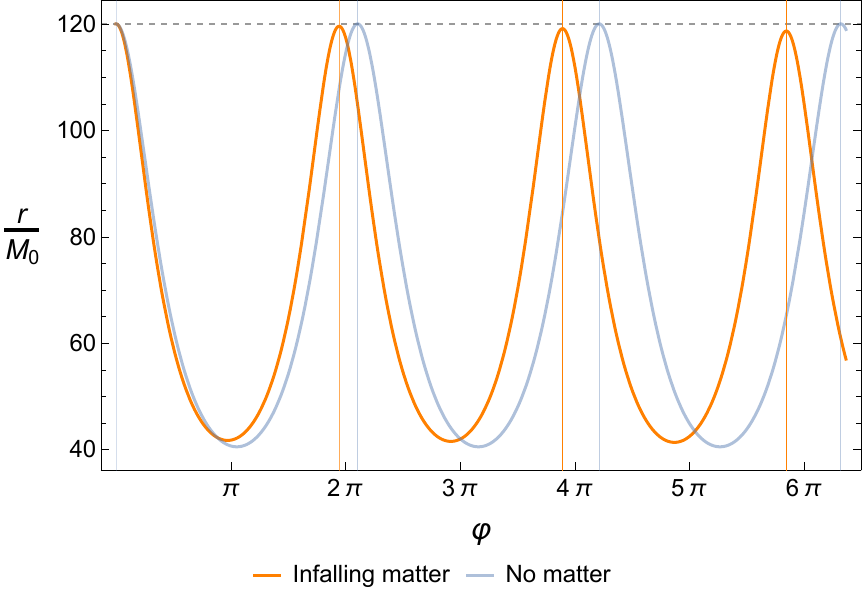}
        \label{fig:shiftC2}
        }
    \vspace{0.5cm} 
    \subfloat[Case 3: \(w=-3/4\)]{
        \includegraphics[width=0.48\linewidth]{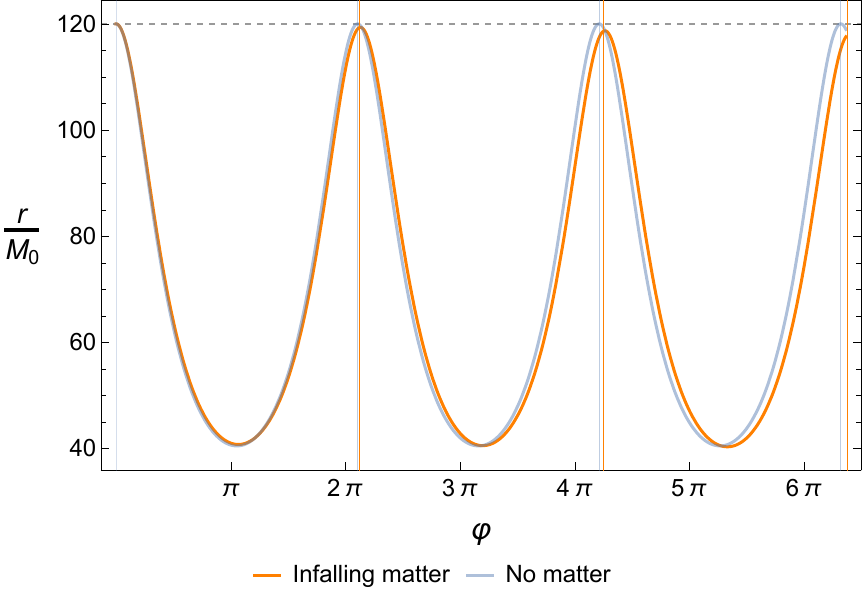}
        \label{fig:shiftC3}
        }
    \hfill
    \subfloat[Case 4: \(w=-4/3\)]{
        \includegraphics[width=0.48\linewidth]{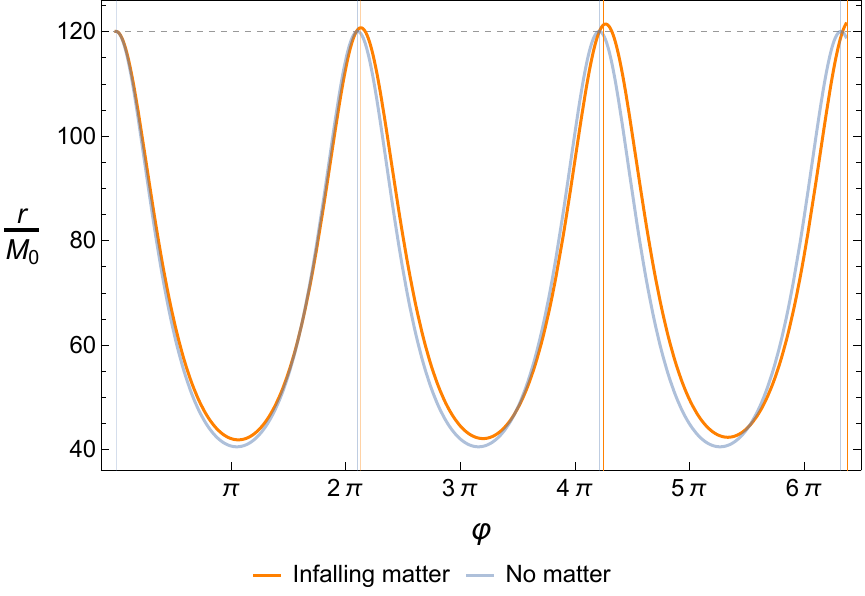}
        \label{fig:shiftC4}
        }
    \caption{Comparison of perturbed and unperturbed orbits \(r(\varphi)\) with apsis shifts calculated using the OOE method for different values of $w$.}
    \label{fig:groupB}
\end{figure}

In Fig.~\ref{fig:groupB}, we plot the 
radius $r$ as a function of $\varphi$ along the geodesics motion. 
We also show the trajectory in the Schwarzschild spacetime as a reference. The vertical lines indicate the longitudes of the apocenter predicted by the method of osculating orbital elements. 
Since the orbital elements change in time, the values of $\Delta \omega_0$, $\Delta \omega_Q$ and $\Delta\omega_\rho$ also change. 
As stated below Eq.~\eqref{eq:d2rdtau}, we can estimate 
those values by using the values of $M_n$, $e_n$ and $a_n$ at 
$n$-th
cycle instead of the initially set values of $M_0$, $e$ and $a$. 
To accurately track the evolution of the orbital parameters across multiple revolutions, we apply the OOE method iteratively. Each orbital pseudocycle, is treated as an oscillation characterized by a maximum radial distance ($r_n^{\rm max}$) and a minimum radial distance ($r_n^{\rm min}$). 
Specifically, 
these extrema define 
the simi-major axis $a_n$ and the eccentricity $e_n$ as 
\begin{equation}
    a_n = \frac{r_n^{\rm min} + r_n^{\rm max}} {2}
\end{equation}
and 
\begin{equation}
    e_n = \frac{r_n^{\rm max} - r_n^{\rm min}} {r_n^{\rm max} + r_n^{\rm min}}, 
\end{equation}
respectively. 
Since the system involves an accreting black hole, the mass of the system evolves. To account for this, we define an effective mass at the beginning of each cycle 
as follows: 
\begin{equation}
    M_n = \frac{l^2}{a_n(1-e_n^2)}. 
\end{equation}

\begin{figure}[ht!]
    \centering
    \subfloat[Case 1: \(w=1/3\)]{
        \includegraphics[width=0.48\linewidth]{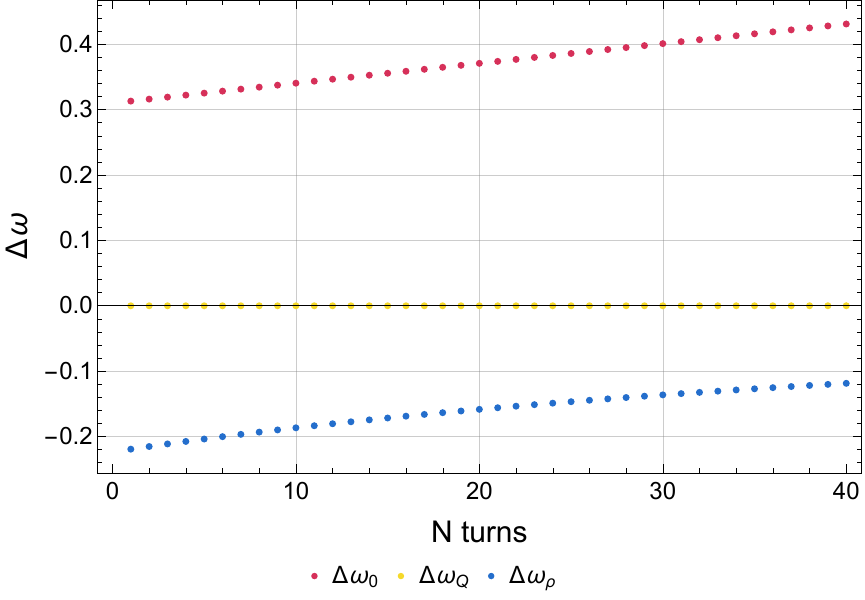}
        \label{fig:components1}
        }
    \hfill
    \subfloat[Case 2: \(w=2/3\)]{
        \includegraphics[width=0.48\linewidth]{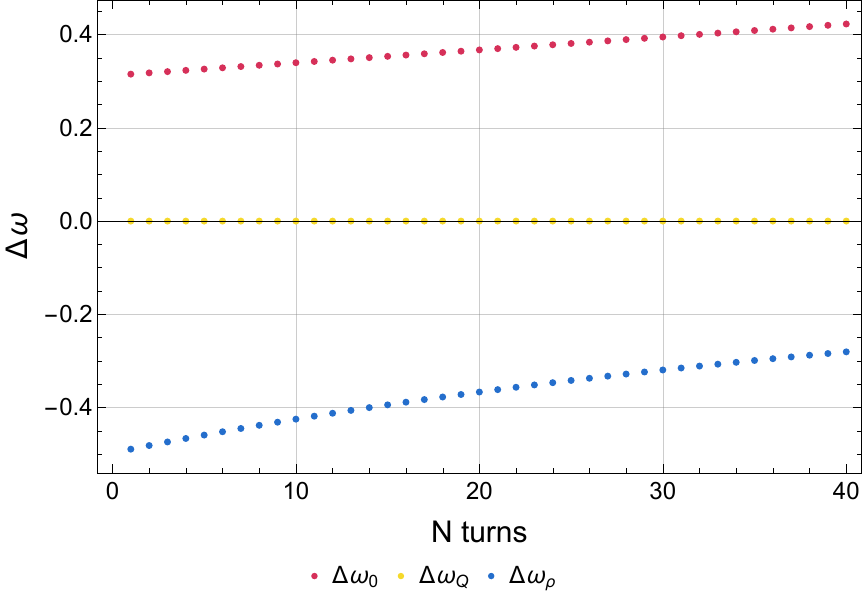}
        \label{fig:components2}
        }
    \vspace{0.5cm} 
    \subfloat[Case 3: \(w=-3/4\)]{
        \includegraphics[width=0.48\linewidth]{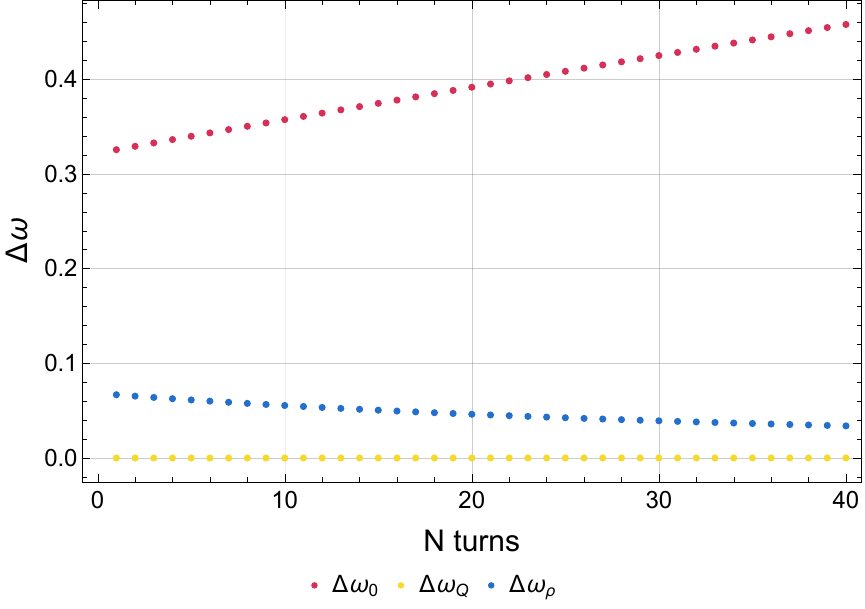}
        \label{fig:components3}
        }
    \hfill
    \subfloat[Case 4: \(w=-4/3\)]{
        \includegraphics[width=0.48\linewidth]{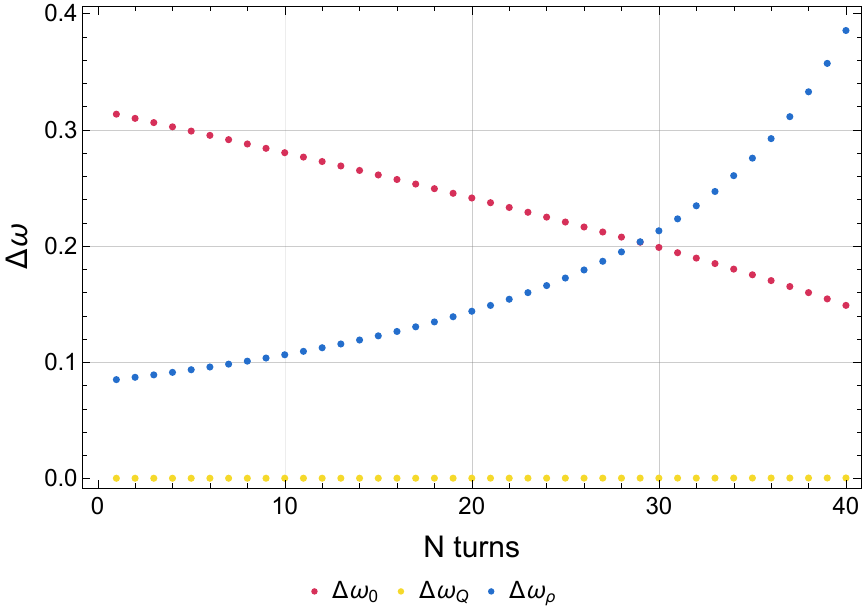}
        \label{fig:components4}
        }
    \caption{Apsis angle shift per orbital cycle components \(\delta \omega (N)\) computed using the OOE method for different values of $w$.}
    \label{fig:groupC}
\end{figure}

The evolution of the values of $\Delta \omega_0$, $\Delta \omega_Q$ and $\Delta\omega_\rho$ are shown in Fig.~\ref{fig:groupC}. 
Fig.~\ref{fig:groupC} shows how the apsis shift varies across the four cases. 
First, the contribution of $\Delta \omega_Q$ is always negligibly smaller than other components. 
In cases 1 and 2, the sign of $\Delta \omega_\rho$ is 
negative,
which means that the total shift is 
smaller
than in the vacuum case, with pressure differences leading to 
variations in the significance of $\Delta\omega_\rho$. 
For case 1, the absolute value of $\Delta \omega_0$ is larger than $|\Delta\omega_{\rho}|$, and $\Delta \omega_{\rm total}$ is positive, indicating apsis shift advance in total. 
In contrast, for case 2, we can see that the initial magnitude of $|\Delta\omega_{\rho}|$ is 
larger 
than $\Delta\omega_0$, leading to a 
periapsis retreat 
at the beginning. 
As the orbit shrinks in time, the value of $\Delta \omega_0$ increases and $|\Delta \omega_\rho|$ decreases. Then, for case 2, as is also mentioned in Sec.~\ref{sec:style}, the periapsis shift turns to be prograde after a while from the initial retrograde shift. 
In cases 3 and 4, 
as can be seen Table~\ref{table:omegas} or  Figs.~\ref{fig:shiftC3}, \ref{fig:shiftC4}, \ref{fig:components3} and \ref{fig:components4}, the sign of $\Delta\omega_\rho$ is positive consistently to the expression \eqref{eq:delomrhoappr}, and therefore the value of 
the total periapsis advance 
is initially larger than the vacuum case.




\section{Effects on the redshift of the orbiting test particle}
\label{sec:redshift}

Let us discuss possible observational consequences of the perturbed particle trajectories. One of the useful observables is the redshift between orbiting particles and the distant observer.  
In general, 
the expression for redshift $Z$ is given by
\begin{equation}
    Z = \frac{p^{\mu} k_{\mu}^{\rm em}}{U^{\mu} k_{\mu}^{\rm obs}}-1, 
\end{equation}
where $p^{\mu}$ and $U^\mu$ are the four-velocities of the test particle and the observer, respectively, and $k_\mu^{\rm em}$ and $k_\mu^{\rm obs}$ are the four-momentum of the photon at the emission and observed points, respectively. 
First, we take the coordinates such that the test particle is orbiting on the $x$-$y$ plane with the coordinates $x:=r\sin\theta\cos\phi$, $y:=r\sin\theta\sin\phi$ and $z:=r\cos\theta$. 
In reality, the trajectories of photons traveling from the source to the observer are 
non-trivial due to the gravitational lensing effects, and a numerical procedure to find such a trajectory is needed. 
Since our current purpose is not to precisely calculate the value of the redshift, but to demonstrate the possible observational effects, we approximate the photon trajectories to be along the $x$-direction. We also assume the observer is at rest for the coordinate frame, that is, $U^\mu=(1, 0, 0, 0)$.

If the vector field $\partial/\partial V$ is a Killing vector field like in the background Schwarzschild metric, since the $V$ component of the photon four-momentum $k_V$
is conserved along the null geodesic, the redshift is calculated as 
\begin{equation}
    Z = \frac{p^{\mu}k^{\rm em}_{\mu}}{k^{\rm em}_V} -1 = p^V +p^x \frac{k^{\rm em}_x}{k^{\rm em}_V} +p^y \frac{k^{\rm em}_y}{k^{\rm em}_V} - 1 .
    \label{eq:perturbatedZ}
\end{equation}
Since this expression can be evaluated at the point of the particle without solving the null geodesic equation for the photon, although it is not exactly consistent with the non-stationary background spacetime, we adopt this approximate expression for convenience. 
That is, we do not take the effects of the non-stationary character of the spacetime on the photon motion into account in the following analyses. 

The components of the particle four-momentum $p^V$, $p^x$ and $p^y$ can be calculated from the numerical integration of the geodesic equations. 
The contravariant four-momentum of the photon $k^\mu_{\rm em}:=g^{\mu\nu}k^{\rm em}_{\nu}$ has only $V$ and $x$ components under our assumption. 
From the null condition $k^\mu_{\rm em}k_\mu^{\rm em}=0$, we can write the four-momentum $k^\mu_{\rm em}$ as 
\begin{equation}
    k^{\mu}\simeq\left(a, 
    a(1+\lambda)\frac{r}{y^2}\left(-x+\sqrt{x^2+\left(1-\frac{2M}{r}\right)y^2}\right)
    ,0,0 \right).
\end{equation}
where $a$ is an arbitrary normalization constant, which does not affect the value of the redshift \eqref{eq:perturbatedZ}. 
Then the covariant components $k^{\rm em}_x$ and $k^{\rm em}_y$ can be straightforwardly calculated.

\begin{figure}[htbp]
    \centering
    \includegraphics[width=100mm]{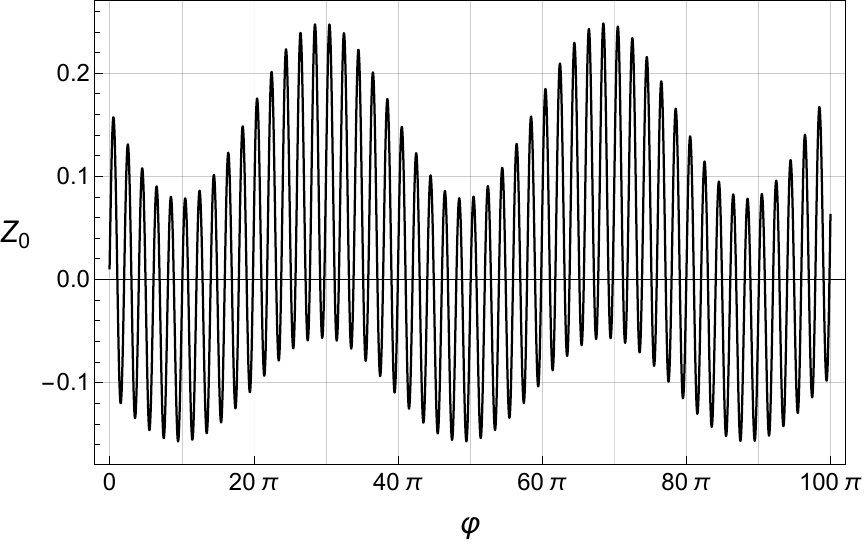}
    \caption{Redshift distribution in the orbital plane of a test particle under the Schwarzschild metric, used as a reference to compare with perturbed solutions.}
    \label{fig:redshiftconf}
\end{figure}

Figure \ref{fig:redshiftconf} depicts periodic redshift fluctuations in the vacuum scenario with 
rapid
oscillations superimposed on a 
slower
cycle.

\begin{figure}[ht!]
    \centering
    \subfloat[Case 1: \(w=1/3\)]{
        \includegraphics[width=0.48\linewidth]{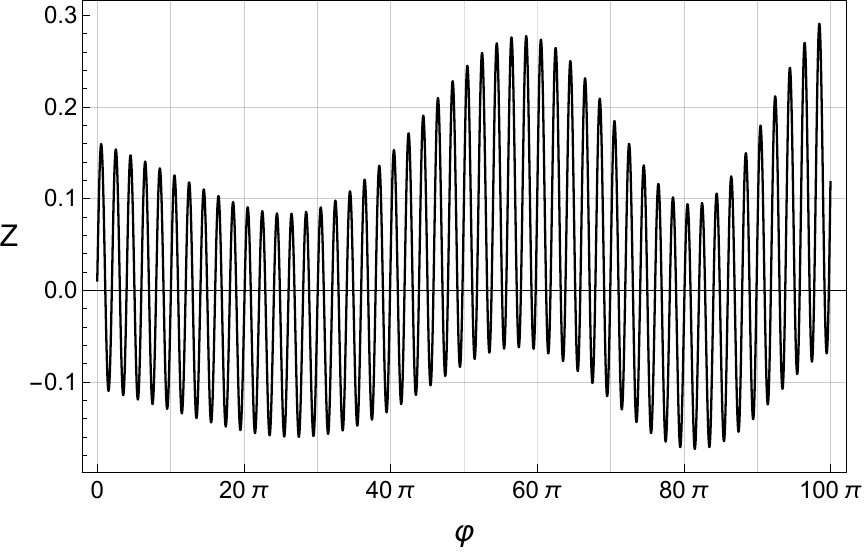}
        \label{fig:red1}
        }
   \hfill
    \subfloat[Case 2: \(w=2/3\)]{
        \includegraphics[width=0.48\linewidth]{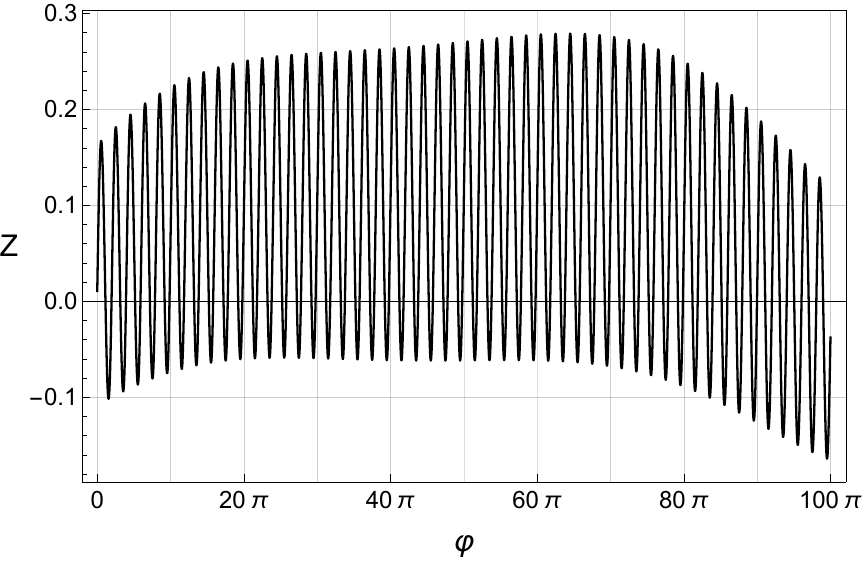}
        \label{fig:red2}
        }
    \vspace{0.5cm} 
    \subfloat[Case 3: \(w=-3/4\)]{
        \includegraphics[width=0.48\linewidth]{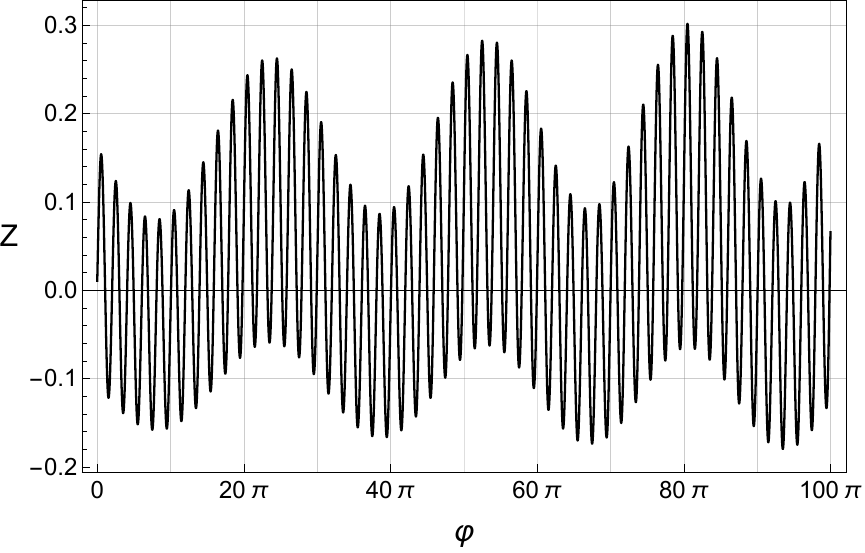}
        \label{fig:red3}
        }
    \hfill
    \subfloat[Case 4: \(w=-4/3\)]{
        \includegraphics[width=0.48\linewidth]{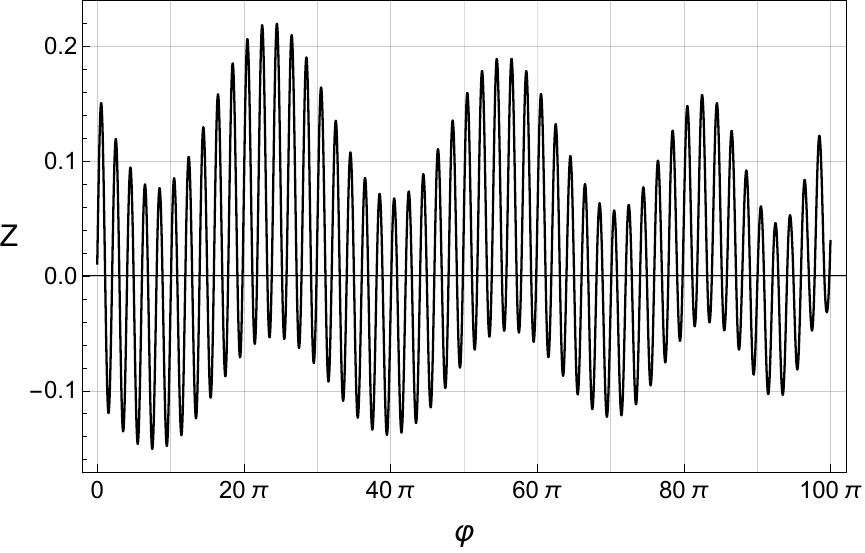}
        \label{fig:red4}
        }
    \caption{Redshift distribution \(Z(\varphi)\) for different values of $w$ in the perturbed spacetime scenario.}
    \label{fig:groupD}
\end{figure}
Notably, the maximum redshift ($Z>0$) is consistently larger than the maximum blueshift ($Z<0$). 
%
This contribution 
can be attributed to the gravitational redshift caused by the black hole’s gravitational field, which is always positive overall. 
%
The shorter period 
oscillations are caused by the relativistic Doppler effect, resulting from changes in the direction of the particle's velocity. 
Superimposed on this, we observe a 
longer 
oscillatory pattern that arises due to the precession of the apsidal angle around the black hole.
%
Under the current setting, the 
maximum/minimum corresponds to the point at which the $x$-component of the test particle's velocity is negatively/positively greatest. 
Motion purely towards the observer ($p^y=0$, $p^x>0$) produces the peak blueshift ($Z<0$), and this effect is enhanced as the particle nears the black hole 
due to the faster motion.
Conversely, motion along the negative $x$-axis, away from the observer, produces a peak redshift ($Z>0$), which is similarly maximized when the test particle is closest to the black hole.

The redshift patterns vary significantly depending on the fluid type and corresponding $w$ parameter. For regular matter 
($w>-1$ and $Q>0$)
the amplitude of the redshift modulation 
increases as 
the particle orbit shrinks. 
In the exotic and phantom energy cases 
($w<-1$ and $Q<0$)
the particle orbit expands with the negative energy flux into the black hole, and the amplitude of the redshift modulation decreases in time
as seen in Figure \ref{fig:red4}.

\section{Conclusions and summary}
\label{sec:sumcon}

In this paper, we have demonstrated the effects of the stationary radial inflow of a matter field into the spherically symmetric black hole on the orbital motion of a massive test particle. Specifically, we have considered the perfect fluid accretion with 
the linear equation of state $p=w\rho$, where $p$, $\rho$ and $w$ are the energy density, pressure and the constant characterizing the equation of state. 
We constructed the background spacetime with 
a perturbative approach to analyze the effects of fluid accretion on black hole spacetime curvature, offering a flexible model that incorporates mass accretion. 
More concretely, we derived the first-order metric 
of the perturbed spacetime due to the existence of the 
accreting matter on the Schwarzschild solution. 
This approach enabled us to explore the dynamics of test particles orbiting black holes 
dressed with different fluid types, characterized by various equations of state ($w$). These effects significantly shape the particle orbital 
motion including the 
apsis shift. 

As for the fluid equation of state, we also accepted exotic ones. 
From a phenomenological point of view, the modification due to the fluid accretion 
could be effectively caused by a gravity theory modification, as mentioned in the introduction. In this case, we may not have any reason to stick to regular fluid satisfying energy conditions. Therefore we took the area of our investigation wider with an open mind. 

We have demonstrated that the difference in the particle orbits 
may be probed by using the redshift observation of stars orbiting around the black hole. 
Our redshift analysis reveals that accreted fluid 
increases the 
redshift modulation amplitude for positive fluid accretion ($Q > 0$) and 
reduces
it for negative accretion ($Q < 0$). 
These findings align with theoretical predictions and 
elucidate how accretion 
affects the observation of the black hole vicinity.  


The comparison between perturbed and vacuum scenarios emphasizes the importance of including possible environmental effects. 
These insights contribute to understanding the complex interactions governing accretion processes and their influence on black hole dynamics. 
While our model assumes spherical symmetry, 
it lays a foundation for future studies 
to provide a more comprehensive description of accretion dynamics. Extending this work could enhance simulations of supermassive black holes, such as Sagittarius A* and M87*, with applications to observational data from instruments like the Event Horizon Telescope.

In conclusion, our framework offers a detailed understanding of fluid accretion’s impact on black hole metrics, particle orbits and those observable properties, aligning with and expanding upon established theoretical models. 
Future research will address the outlined limitations and explore more complex scenarios
to advance our knowledge of black hole systems.

\begin{acknowledgments}
    We acknowledge Hiroki Matsuda for his contribution in the initial stage of this work. 
    This work was supported in part by JSPS KAKENHI Grant Numbers 20H05850~(CY), 20H05853~(CY), JP21K20367~(YK), JP23H01170~(YK), 24K07027~(CY) and 25K07281~(CY). 
\end{acknowledgments}
    

\end{document}